\documentclass[onecolumn, letterpaper, 12pt]{IEEEtran}
\usepackage{graphicx,amssymb,epsfig,amssymb,amsmath,setspace,url, cite}

\begin{document}
\title{Optimal Relay-Subset Selection and Time-Allocation in Decode-and-Forward Cooperative Networks}
\author{
  Elzbieta Beres and Raviraj Adve \\
   Dept. of Elec. and Comp. Eng. \\
   University of Toronto \\
   10 King's College Road \\
   Toronto, ON M5S 3G4, Canada \\
    email: ela.beres@utoronto.ca, rsadve@comm.utoronto.ca \\
} \maketitle

\begin{abstract}
We present the optimal relay-subset selection and
transmission-time for a decode-and-forward, half-duplex
cooperative network of arbitrary size. The resource allocation is
obtained by maximizing over the rates obtained for each possible
subset of active relays, and the unique time allocation for each
set can be obtained by solving a linear system of equations. We
also present a simple recursive algorithm for the optimization
problem which reduces the computational load of finding the
required matrix inverses, and reduces the number of required
iterations. Our results, in terms of outage rate, confirm the
benefit of adding potential relays to a small network and the
diminishing marginal returns for a larger network. We also show
that optimizing over the channel resources ensures that more
relays are active over a larger SNR range, and that linear network
constellations significantly outperform grid constellations.
Through simulations, the optimization is shown to be robust to
node numbering.

\end{abstract}

\doublespacing

\section{Introduction} \label{introduction}

Cooperation has become a popular technique to implement diversity
in the absence of multiple antennas at receiving and transmitting
nodes~\cite{SendonarisErkip, LanemanWornell, BeresAdve-TWC}. In
this context, resource allocation in cooperative networks has
recently become an active research area, and has been investigated
under many scenarios and metrics. In this paper, we address the
problem of resource allocation, in terms of channel resources
(time or bandwidth), in multi-relay networks with arbitrary
connections. We describe the contributions of the paper in detail
after a brief review of the pertinent literature.

For the single-relay case, several works have dealt with various
aspects of resource allocation, in terms of power and/or bandwidth
and time. Yao et al. determine the optimal power and time
allocation for relayed transmissions specifically in the low-power
regime~\cite{YaoCaiGiannakis}. Larsson and Cao present various
strategies for allocating power and channel resources under energy
constraints~\cite{LarssonCao}. For the channel resource allocation
problem, however, the authors consider selection combining only
and do not address the scenario of joint decoding of the source
and relay signals. The works
in~\cite{GunduzErkip,YangGunduzBrownErkip, Host-MadsenZhang_Power}
address the problem of power and channel resource allocation under
sum average power constraints. Optimal time and bandwidth
allocation using instantaneous and average channel conditions is
obtained using power control in~\cite{XieZhang}. Channel resource
allocation under fixed power is developed
in~\cite{NingHuiShashaPing}

In networks with multiple relays, the available literature can be
classified into two groups: networks where relays do not
communicate with one another (parallel-relay networks), and
networks without restrictions on relay communication
(arbitrarily-connected networks). Resource allocation for the
former has been addressed in ~\cite{IbrahimiLiang-ICC,
AnghelKavehLuo_Power, AnghelKavehLuo-Orthogonal,
LiangVeeravalliPoor-MaxMin}. Ibrahimi and Liang develop the
optimal power allocation for a multi-relay cooperative OFDMA
amplify-and-forward (AF) system~\cite{IbrahimiLiang-ICC}. By
maximizing the channel mutual information, Anghel et al. find the
optimal power allocation for multiple parallel relays in AF
networks~\cite{AnghelKavehLuo_Power, AnghelKavehLuo-Orthogonal}. A
more general solution is given
in~\cite{LiangVeeravalliPoor-MaxMin} where the authors give the
optimal power and channel resource allocation for a parallel-relay
network with individual power constraints on the nodes.

To the best of our knowledge, the problem of channel resource
allocation for arbitrarily-connected networks and dedicated multiple
access has not been addressed in the literature. In general, works
in the area of multi-relay systems with arbitrary links generally
neglect the bandwidth penalty arising from multiple hops by assuming
either full-duplex nodes, a bandwidth-unconstrained system, or the
availability of channel phase information at the
transmitter~\cite{BoyerFalconerYanikomeroglu-Transactions-2004,
SadekSuLiu, YangHøst-Madsen, ZhangLok, ChenMitraKrishnamachari,
FangHuiPingNing, KhandaniAbounadiModianoZhang, LiLippmanWu,
YuanHeChen, YuanChenKwon, SrinivasanNuggehalliChiasseriniRao,
OngMotani-SECON, AksuErcetin}.

These assumptions, however, are not realistic for practical
wireless networks, where nodes are likely to be half-duplex, phase
information is very difficult to obtain at the transmitter, and
bandwidth is a scarce resource. To fill this void, in this paper
we investigate the problem of resource allocation in a
bandwidth-constrained, cooperative, decode-and-forward (DF),
wireless network, and consider the most general setting where
multiple relays can transmit can cooperate with each other in
transmitting information between source and destination. In this
setting, we address the joint problem of optimal selection of a
relaying subset and allocation of time resources to the selected
relays. The resource allocation is framed in the context of mesh
networks of relatively simple and inexpensive nodes. We
concentrate on resource allocation in terms of transmission time
only, removing power allocation from the optimization; we further
simply the problem by considering orthogonal transmissions. This
is motivated by the need to reduce complexity, allowing for nodes
which can implement the resource allocation simply by switching on
and off. For a system without power allocation, a solution to this
problem provides an upper bound on cooperative performance in
multi-relay network where dedicated channels are assigned for each
source transmission.

To the best of our knowledge, no other work provides a solution to
time-allocation for an arbitrarily connected cooperative network.
The solution can be interpreted as a generalization of the
opportunistic protocol presented by Gunduz and Erkip, where the
relay is active only when it increases the outage
rate~\cite{GunduzErkip}. In terms of the resource allocation
solution, it is also a generalization of the solution
in~\cite{NingHuiShashaPing}, where channel resource allocation is
determined under fixed power for a three-node DF network. The
solution can also be interpreted as a generalization of node
selection~\cite{BeresAdve-TWC, ZhaoAdveLim_SelectionAFPowerAlloc,
BletsasKhistiReedLippman, BletsasShinWin} under relaxed
transmission constraints, where transmission can occur on multiple
time-slots and relays can communicate with one another.

This paper is structured as follows.
Section~\ref{sec:System-Model} describes the system model. In
Section~\ref{sec:Resource-Allocation}
and~\ref{sec:Ch3-Implementation} we develop the proposed resource
allocation scheme and present a significantly simplified recursive
implementation. Simulation results are presented in
Section~\ref{sec:Ch3-Simulations} and concluding remarks
are presented in Section~\ref{sec:Conclusions}. 
\section{System Model} \label{sec:System-Model}
We consider a mesh network of static nodes comprising a source and
destination node and $N$ potential relays. The inter-node channel
powers are denoted as $|a_{ij}|^2$, where $i$ and $j$ represent
the source node $s$, relay nodes $r_k, \: k = 1 \ldots N$, or the
destination node $d$. They are assumed independent of each other
and are modelled as flat, slowly-fading and exponential with
parameter $\lambda$. $\lambda$ is inversely proportional to the
average channel power and is a function of inter-node distance,
$d_{ij}$, through the path loss exponent $p_a$, e.g.,
$1/\lambda_{sd} = (1/d_{sd}^{p_a})$, and $ 1/\lambda_{r_k d} =
(1/d_{r_kd}^{p_a})$. We do not include shadowing into the fading
model, although this can easily be incorporated on an
instantaneous basis. Because the nodes are static, the channels
are assumed to change very slowly with time; we thus assume
knowledge of all channel gains (although \emph{not} channel
phases) at a centralizing unit. This knowledge is essential to our
resource allocation scheme.
%
%
%
%

\begin{figure}
\begin{center}
\includegraphics[width=4.0in]{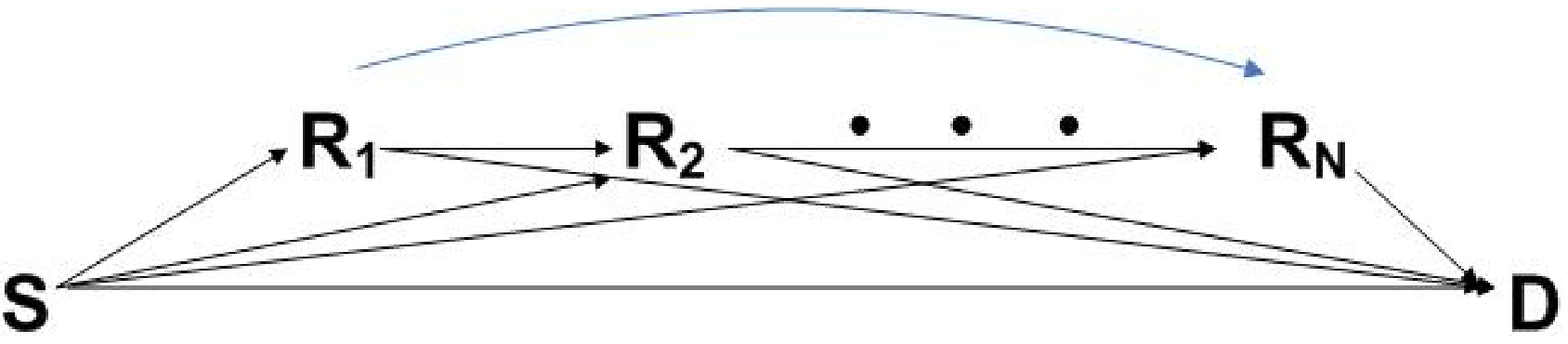}
\caption{Location of the relays with respect to the source and
destination.} \label{fig:Ch3-Relays-Line}
\end{center}
\end{figure}

%
%
With the aim of designing simple and cheap nodes, we assume
half-duplex channels and orthogonal transmissions, which greatly
simplifies receiver design. The relays are assumed to be numbered
in some convenient order such that relay $r_j$  transmits after
$r_i$ if $j>i$. For example, the relays may be in a linear
constellation as shown in Figure~\ref{fig:Ch3-Relays-Line}. We
also assume the DF cooperation strategy with independent
codebooks, which allow for the optimization of system resources
(see~\cite{ChakrabartiErkipSabharwalAazhang} for an overview on
current coding methods for nodes using DF). Note that repetition
coding does not allow for this resource allocation.

With these assumptions, the cooperation framework for the
$N$-relay fully-connected network is as follows. The half-duplex
constraint precludes the relays from transmitting and receiving
simultaneously on the same channel, and the unavailability of
forward-channel phase information at the nodes precludes the nodes
from simultaneous transmissions. The transmission between the
source and destination is thus divided into $N+1$ time-slots, of
duration $t_0$, $t_1$, $\ldots$, $t_{N}$, with $t_0 + t_1 + \ldots
+ t_{N} = 1$. In the first time-slot, the source transmits its
information to all the nodes. The first relay, $r_1$, decodes this
information and the remaining $N$ relays and the destination store
the information for future processing. In the second slot, of
duration $t_1$, the first relay re-transmits the information using
an independent codebook, the second relay decodes the information
from the first relay and the source, and the remaining $N-1$
relays and the destination store the information for further
processing. In general, each relay $r_k$ decodes information from
the source and from the previous relays $r_1 \ldots r_{k-1}$ up to
and including time-slot $t_{k-1}$. This process continues until
all relays have transmitted and the destination attempts to decode
the information.

Assuming that each node uses power $P$ and $W$ Hz per transmission
(noting that although each node transmits for a different amount
of time, the symbol durations and thus the corresponding bandwidth
used by each node is the same), the signal to noise ratio (SNR) at
node $j$ resulting from transmission from node $i$ can be written
as
%
$\mathrm{SNR}_{ij} = \frac{P}{N_0 W} |a_{ij}|^2$,
%
%
where $N_0$ is the noise spectral density. In the rest of the
paper, we use the short-hand notation $L_{ij}$ to denote $\log_2(1
+ \mathrm{SNR}_{ij})$, the capacity of the corresponding channel.
\section{Optimal Resource Allocation and Relay Selection }
\label{sec:Resource-Allocation}
In this section, we solve the joint problem of resource allocation
and relay selection for the network discussed above. Essentially, we
give the optimum values of $t_i$, $i = 0 \ldots N$, such that the
achievable rate between source and destination is maximized. We
begin here with a fully-connected network, where each node is linked
to all other nodes through a non-zero channel.
\subsection{Fully Connected Network}
\label{sec:Resource-Allocation-Fully-Connected}
Consider a source-destination pair communicating with the help of
$N$ relays. Assuming that each relay is active, the mutual
information at each relay and destination can be written as
\begin{eqnarray}
I_1(t_0) &=&  t_0 L_{s r_1} \label{eq:I-1}, \\
I_{k}(t_0, t_1, t_2, \ldots, t_{k-1}) &=&  t_0 L_{sr_k} +  t_1
L_{r_1 r_k} + \ldots +  t_{k-1} L_{r_{k-1} r_k} \label{eq:I-k},
\\
\nonumber I_{D}(t_0, t_1, t_2, \ldots, t_{k}, \ldots, t_{N-1},
t_{N}) &=& t_0 L_{sd} + t_1 L_{r_1 d}  + \ldots +  t_{k-1}
L_{r_{k-1} d} + \ldots + t_{N}  L_{r_{N} d}, \\ \label{eq:I-D}
\end{eqnarray}
\noindent where $I_k$ and $I_D$ denote the mutual information at
relay $r_k$ and the destination, respectively.

With all $N$ relays cooperating, the maximum achievable rate under
orthogonal transmissions is the minimum of the mutual information
obtained at each individual relay node:
\begin{eqnarray}
\label{eq:R-0}  \nonumber R_N = \max_{t_0, \ldots, t_{N}} &\min&
\{ I_1(t_0), I_2(t_0, t_1), \ldots, I_{k-1}(t_0, \ldots, t_{k-2}),
I_{k}(t_0, \ldots, t_{k-1}),  I_{k+1}(t_0, \ldots, t_{k}),
\\
 && \ldots, I_{N}(t_0, \ldots, t_{k}, \ldots, t_{N-1}),
I_{D}(t_0, \ldots, t_{k},
\ldots, t_{N-1}, t_{N}) \}, \\
\nonumber  \mathrm{such} \: \mathrm{that}  && t_i \geq 0, \quad
\forall
i, \\
\nonumber  && t_0 + t_1  + \ldots t_{N} \leq 1.
\end{eqnarray}
The above expression is a straightforward generalization of the
cut-set bound for the single-relay network. This generalization
maintains orthogonal transmissions for each relay, a model which
represents practical networks with simple nodes that cannot
implement complex interference cancelation. We use this model as
the basis of the optimization in the rest of this paper. We note,
however, that because each relay transmits using an orthogonal
channel, $R_N$ is clearly not the channel capacity. For literature
on the channel capacity of arbitrarily-connected networks, we
direct the reader to~\cite{KramerGastparGupta, XieKumar,
GuptaKumar-LargeNetworks, AvestimehrDiggaviTse,
ReznikKulkarniVerdu, RazaghiYu-Multirelay} for full-duplex relays,
and~\cite{RostFettweis-Multirelay} for half-duplex relays.

For reasons that will soon become clear, let us consider the case
with relay $r_k$ removed from the network. The maximum achievable
rate $R_{N-1}^k$ becomes
\begin{eqnarray}
\label{eq:R-k}  R_{N-1}^k = \max_{t_0, \ldots, t_{k-1}, t_{k+1},
\ldots t_{N}} &\min&  \{
I_1(t_0) \ldots, I_{k-1}(t_0, \ldots, t_{k-2}), I_{k+1}(t_0,
 \ldots, t_{k-1}),
 \\ \nonumber &&  \ldots, I_{D}(t_0, \ldots, t_{k-1}, t_{k+1}, \ldots , t_{N})
 \}, \\ \nonumber
 \qquad \qquad  \mathrm{such} \: \mathrm{that}  \quad && t_i \geq 0, \quad \forall
 i,
 \\ \nonumber && t_0 + \ldots t_{k-1} + t_{k+1} + \ldots t_{N} \leq 1.
\end{eqnarray}

\noindent Removing relay $r_k$ is thus equivalent to removing
$t_{k}$ and $I_k$ from the optimization. [We use the subscript in
$R_{N-1}^k$ to denote the maximum number of \emph{potentially}
active relays, and the superscript to denote the relay removed].
The maximum rate at which the source can transmit to the
destination can thus be written as the maximum of the rate
obtained by using all $N$ relays, and the rate obtained by
successively removing each relay:
\begin{eqnarray}
R_T = \max \{  R_N, R_{N-1}^1, R_{N-1}^2, \ldots, R_{N-1}^N \}.
\label{eq:R-T}
\end{eqnarray}
\noindent If $R_T =  R_{N-1}^k$, the maximum rate can be obtained
by iterating through~(\ref{eq:R-0}) and~(\ref{eq:R-k}),
successively removing a relay each step. Note that obtaining
$R_{N-1}^k$ includes the cases where two or more relays are
removed. In theory, therefore, all $2^N$ possible cases must be
checked.

Let $(t_0^*, t_1^*, \ldots, t_N^*)$ denote the resource allocation
that solves the optimization problem. We begin an outline of the
solution to the optimization problem in~(\ref{eq:R-0}),
(\ref{eq:R-k}) and~(\ref{eq:R-T}) with the following proposition.

{\bf Proposition 1:} With a maximum number of potential relays $N$,
the maximum achievable rate  $R_T = R_{N}$ only if $t_k^* \neq 0$,
$\forall k$. Otherwise, if $t_k^* = 0$, $R_T = R_{N-1}^k$.

{\bf Proof:} With exactly $N$ active relays, and with $k< n < N$,
the resulting rate can be written explicitly as:
\begin{eqnarray}
 \nonumber  R_{N} =   \max_{t_0, \ldots, t_{N}} &\min&  \{
(t_0 L_{sr_1}), (t_0 L_{sr_2} + t_1 L_{r_1 r_2}), \ldots ,
(t_0 L_{sr_k}  + \ldots + t_{k-1} L_{r_{k-1} r_k}), \\
\nonumber && (t_0 L_{sr_n} + \ldots + t_{k-1} L_{r_{k-1} r_n} +
t_{k} L_{r_{k} r_n} +  t_{k+1} L_{r_{k+1} r_n} \ldots t_{n-1}
L_{r_{n-1}
r_n}), \ldots, \\
&& ( t_0 L_{sd}  + \ldots + t_{k-1} L_{r_{k-1} r_d} + t_{k} L_{r_{k}
r_d} + t_{k+1} L_{r_{k+1} r_d} + \ldots t_{N} L_{r_{N} r_d} )
\}.
\end{eqnarray}

Setting $t_k = 0$ gives
\begin{eqnarray}  \label{eq:removing-rk}
 \nonumber  R_{N} =   \max_{t_0, \ldots, t_{k-1}, t_{k+1}, \ldots t_{N}} &\min&  \{
(t_0 L_{sr_1}), (t_0 L_{sr_2} + t_1 L_{r_1 r_2}), \ldots ,
(t_0 L_{sr_k}  + \ldots + t_{k-1} L_{r_{k-1} r_k}), \\
\nonumber && (t_0 L_{sr_n} + \ldots + t_{k-1} L_{r_{k-1} r_n} +
t_{k+1}
L_{r_{k+1} r_n} \ldots t_{n-1} L_{r_{n-1} r_n}), \ldots, \\
&& ( t_0 L_{sd}  + \ldots + t_{k-1} L_{r_{k-1} r_d}  + t_{k+1}
L_{r_{k+1} r_d} + \ldots t_{N} L_{r_{N} r_d} )
\label{Ch3-Eq-tk-Full}
\} \\
\nonumber \leq
\max_{t_0, \ldots, t_{k-1}, t_{k+1}, \ldots t_{N}} &\min&  \{
(t_0 L_{sr_1}), (t_0 L_{sr_2} + t_1 L_{r_1 r_2}), \ldots ,
\\
\nonumber && (t_0 L_{sr_n} + \ldots + t_{k-1} L_{r_{k-1} r_n} +
t_{k+1}
L_{r_{k+1} r_n} \ldots t_{n-1} L_{r_{n-1} r_n}), \ldots, \\
\label{Ch3-Eq-tk-Zero}
&& ( t_0 L_{sd}  + \ldots + t_{k-1} L_{r_{k-1} r_d}  + t_{k+1}
L_{r_{k+1} r_d} + \ldots t_{N} L_{r_{N} r_d} )\} \\
\label{eq:R-N-1-k}
 && =      R_{N-1}^{k-1} ,
\end{eqnarray}
\noindent since~(\ref{Ch3-Eq-tk-Zero}) has one fewer term in the
minimization than~(\ref{Ch3-Eq-tk-Full}). $\blacksquare$

To solve the optimization problem of~(\ref{eq:R-0}) we thus
require only the critical points for which $t_k^* \neq 0, \forall
k$. In the following proposition, we show that for each $R_N$,
i.e., given a set of potential relays, only one solution satisfies
$t_k^* \neq 0, \forall k$.

{\bf Proposition 2:} The unique solution to the optimization
problem in the minimization in~(\ref{eq:R-0}) for which $t_k^*
\neq 0, \forall k$ is given by $I_1(t_1) = I_2(t_1, t_2) = \ldots
= I_N(t_1, \ldots, t_N) = I_D(t_1, \ldots, t_{N})$.

{\bf Proof:} We consider all possible critical points obtained from
the optimization in~(\ref{eq:R-0}). The points are obtained either
by maximizing each individual term in~(\ref{eq:R-0}) or by
intersecting all possible combinations of the terms
in~(\ref{eq:R-0}). We show that the only solution leading to
non-zero solutions results from intersecting every term
in~(\ref{eq:R-0}).

The critical points for the optimization problem can be obtained by
solving the following:

\begin{enumerate}
\item Maximize the individual terms in~(\ref{eq:R-0}) except $I_d(t_0, \ldots,
t_N)$:
\begin{equation}
 \forall k \leq N, \hspace*{0.5in} \max_{t_0,\ldots, t_{k-1}}
I_k(t_0, \ldots t_{k-1})   \quad \mathrm{s.t.} \quad t_0 + \ldots +
t_{k-1} \leq 1.
\end{equation}
Because the optimization is not over $t_m, \forall k \leq m \leq
N$, the solution to this problem clearly has all $t_m = 0, \forall
k \leq m \leq N$, and thus cannot be a solution to the overall
optimization problem.
\item Maximize $I_d(t_0, \ldots,
t_N)$:
\begin{equation}
\max_{t_0,\ldots, t_N}  I_d(t_0, \ldots t_{N})
 = \max_{t_0,\ldots, t_N} \{ t_0 L_{sd}  + \ldots +  t_{N} L_{r_{N} r_d}
 \}, \quad \mathrm{s.t.} \quad t_0 +
\ldots + t_N \leq 1.
\end{equation}
In this case, all variables are included in the optimization. It is
easy to show, however, that this function is maximized by selecting
the largest $L$ value, i.e., evaluating the Kuhn-Tucker conditions
leads to a solution of the form $t_m = 1, t_k = 0, \forall k \neq
m$, where $m = \arg \max_k\{ L_{sd}, L_{r_1 d}, \ldots, L_{r_{k} d},
\ldots, L_{r_N d}  \}$. Therefore, this solution is also not a
solution to the overall optimization problem.
\item
Maximize the function that results from the intersection of all
possible combinations of the functions $I_k$. Let $\mathcal{M}$
denote all possible subsets of $\{1 \ldots N \}$. $\mathcal{M}$
then contains $2^N$ such subsets, i.e., $|\mathcal{M}| = 2^N.$
Consider one such subset $\delta_k = (m_1, m_2, \ldots, m_k )$,
with $m_1 < m_2<  m_k$. One critical point then is
\begin{equation}
\max_{t_0,\ldots, t_{m_{k-1}}}  I_{m_k}(t_0, \ldots t_{m_{k-1}})
\end{equation}
such that
\begin{equation}
I_{m_1}(t_0, \ldots t_{m_1 - 1}) = I_{m_2}(t_0, \ldots t_{m_2-1}) =
\ldots = I_{m_k}(t_0, \ldots t_{m_k - 1}).
\end{equation}
This optimization then gets repeated for all sets $\delta_k \in
\mathcal{M}$. In all but one combination, this optimization is not
over all the variables $\{t_0, \ldots t_N \}$. As in point (1), this
maximization also leads to $t_k = 0$ for some value of $k$.
\item Maximize the intersection of all terms in~(\ref{eq:R-0}):
\begin{equation}
I_1(t_0) = I_2(t_0, t_1) = \ldots = I_N(t_0, \ldots, t_{N-1}) =
I_d(t_0, \ldots, t_N). \label{eq:system-equations}
\end{equation}
This is the only case that leads to $t_k \neq 0, \forall k = 0
\ldots N$. $\blacksquare$
\end{enumerate}
Essentially, this proposition shows that if all $N$ relays are to
contribute, all terms in the minimization in~(\ref{eq:R-0}) must
be equal. This proposition applies to any value of $N$. Therefore,
if the optimal solution has $k < N$ relays, an expression
like~(\ref{eq:R-0}) can be written for those $k$ relays.
\subsection{Optimal solution}
The linear system of equations in~(\ref{eq:system-equations}) has
a simple solution. Setting each equation to a constant, solving
for the vector of unknowns $\mathbf{t} = (t_0 \ldots t_N)$ and
normalizing, we obtain
\begin{eqnarray}
\mathbf{L_{N+1}} \mathbf{t}_{N+1} = \mathbf{1}_{N+1}, \Rightarrow
\mathbf{t}_{N+1} = \frac{\mathbf{L}_{N+1}^{-1}\mathbf{1}_{N+1}}{
||\mathbf{L}_{N+1}^{-1}\mathbf{1}_{N+1}||_1}  =
\frac{\mathbf{L}_{N+1}^{-1}\mathbf{1}_{N+1}}{ \mathbf{1}_{N+1}^T
\mathbf{L}_{N+1}^{-1} \mathbf{1}_{N+1}},
\label{eq:solution}
\end{eqnarray}
\noindent where $ || \mathbf{v} ||_1$ denotes the sum of the
elements of $\mathbf{v}$, i.e., the $1$-norm. $\mathbf{1}_{N+1}$
is the length-$(N+1)$ vector of ones and $\mathbf{L}_{N+1}$ is the
$(N+1) \times (N+1)$ rate matrix
\begin{eqnarray} \label{eq:matrix}
\mathbf{L}_{N+1} =  \left[\begin{array}{ccccc} L_{s r_1} & 0 & 0 & \ldots & 0\\
L_{s r_2} & L_{r_1 r_2} & 0 & \ldots & 0\\
L_{s r_3} & L_{r_1 r_3} &  L_{r_2 r_3}& \ldots & 0\\
\vdots & \vdots &  \vdots & \ddots & 0\\
L_{s d} & L_{r_1 d} &  L_{r_2 d}& \ldots &  L_{r_N d} \\
\end{array} \right].
\end{eqnarray}
The solution in~(\ref{eq:solution}) does not guarantee that the
constraint $t_k > 0 \quad \forall k = 0 \ldots N$ is satisfied. To
ensure that only solutions for which this constraint is satisfied
are considered, we again consider the set $\mathcal{M}$.
%
%
%
Each entry in the set corresponds to a rate matrix,
$\mathbf{L}_{m}$, similar to that in~(\ref{eq:matrix}), formed
using the relays in that entry of the set. Furthermore, let $|m|$
denote the size of the rate matrix $\mathbf{L}_{m}$. A relay set
and its corresponding solution, denoted as $\mathbf{t}_m$, is
included as a potential solution if $\mathbf{t}_m$ satisfies the
constraint, i.e.,
\begin{eqnarray}
\mathbf{t}_m > \mathbf{0}_{|m|},
\end{eqnarray}
\noindent where
$\mathbf{0}_{|m|}$ is the all-zero vector of size $|m|$,
$\mathbf{0}_{|m|} = [0, 0, 0, \ldots 0]^T$ and the inequality
operates on an element-by-element basis. Let the set $\mathcal{K}$
form the subset of $\mathcal{M}$ that comprises all potential
solutions. Let $\mathbf{L}_{k}$, $\mathbf{t}_k$ and $|k|$ denote
the rate matrix, its corresponding solution and size,
respectively, for each entry of the set $\mathcal{K}$. Note that
the number of active relays being considered in each entry is
$|k|$ - 1. Finally, the optimum solution can be obtained by
solving~(\ref{eq:solution}) for all possible combinations of
active relays in the set $\mathcal{K}$ i.e.,
\begin{eqnarray} \label{eq:Ch3-final-solution}
\mathbf{t}^* = \max_\mathcal{K} \frac{\mathbf{L}_k^{-1}
\mathbf{1}_{|k|} }{ \mathbf{1}_{|k|}^T
\mathbf{L}_k^{-1}\mathbf{1}_{|k|}}, \forall k = 1, \ldots,
|\mathcal{K}|.
\end{eqnarray}
\noindent Given that entry $k^*$ corresponds to $\mathbf{t}^*$, the
maximum achievable rate vector can thus be written as
\begin{eqnarray}
 \mathbf{L}_{k^*} \mathbf{t}^* =
 \mathbf{L}_{k^*} \frac{\mathbf{L}_{k^*}^{-1} \mathbf{1}_{|{k^*}|}  }{ \mathbf{1}_{|{k^*}|}^T
\mathbf{L}_{k^*}^{-1}\mathbf{1}_{|{k^*}|}} =
\frac{ \mathbf{1}_{|{k^*}|}  }{ \mathbf{1}_{|{k^*}|}^T
\mathbf{L}_{k^*}^{-1}\mathbf{1}_{|{k^*}|}},
\end{eqnarray}
\noindent and the maximum achievable rate, $R^*$, is
\begin{eqnarray}
R^* = \frac{ 1  }{ \mathbf{1}_{|{k^*}|}^T
\mathbf{L}_{k^*}^{-1}\mathbf{1}_{|{k^*}|}}, \label{eq:Ch3-Rate-Star}
\end{eqnarray}
\noindent Note that the solution described above is equivalent to
the iterative maximization in~(\ref{eq:R-T}), and that removing a
relay $r_k$ translates to removing the $k^\mathrm{th}$ row and
$(k+1)^\mathrm{th}$ column from the rate matrix
in~(\ref{eq:matrix}). Removing the first relay, for example,
reduces the rate matrix in~(\ref{eq:matrix}) to
\begin{eqnarray} \label{eq:partial_matrix1}
\mathbf{L}_{N} =  \left[\begin{array}{cccc}
L_{s r_2}  & 0 & \ldots & 0\\
L_{s r_3} &  L_{r_2 r_3}& \ldots & 0\\
\vdots  &  \vdots & \ddots & 0\\
L_{s d}  &  L_{r_2 d}& \ldots &  L_{r_N d} \\
\end{array} \right].
\end{eqnarray}
\noindent Since $|\mathcal{M}| = 2^N$, $2^N$ possible solutions
must be tested to find the global optimum.
\subsection{Numbering}
\label{Ch3:Numbering}

In Section~\ref{sec:Resource-Allocation-Fully-Connected}, we gave
the solution to the optimization problem for a network with nodes
numbered as in Figure~\ref{fig:Ch3-Relays-Line}. The numbering of
the relay nodes impacts performance through causality: relay $r_k$
decodes information from relay $r_{k-1}$, but not vice-versa. A
complete solution to the optimization problem must therefore take
into account an optimal numbering scheme. In the worst case (in
terms of computational power), an optimal solution can be obtained
for a specific numbering scheme, and the truly optimal solution
can be maximized over all possible numbering schemes.

Clearly, such an approach is impractical. Although a search for an
optimal or effective sub-optimal solution is beyond the scope of
this paper, we study the effects of numbering on the solution and
resulting rate by considering some numbering schemes based on
heuristics. We consider two approaches: numbering based on average
channel conditions, and numbering based on instantaneous channel
conditions.
\subsubsection{Numbering based on average channel conditions}
\begin{figure}
\begin{center}
\includegraphics[width=4.0in]{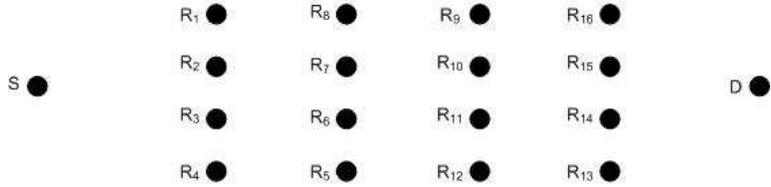}
\caption{ Numbering in a square $4 \times 4$ network in linear order
} \label{fig:Ch3-Numbering-linear}
\end{center}
\end{figure}

\begin{figure}
\begin{center}
\includegraphics[width=4.0in]{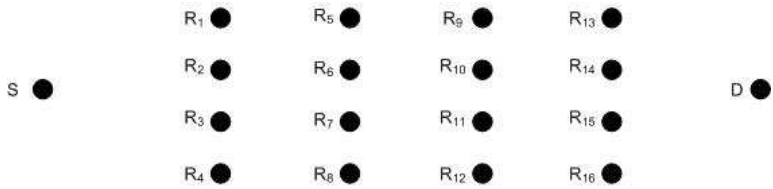}
\caption{Numbering in a square $4 \times 4$ network in descending
order } \label{fig:Ch3-Numbering-descending}
\end{center}
\end{figure}

In the case of the linear network in
Figure~\ref{fig:Ch3-Relays-Line}, the numbering is trivial: node
numbers increase away from the source and towards the destination.
In the case of square network with nodes arranged in a grid, we
consider two numberings which we refer to as \textbf{Average
Descending Numbering} and \textbf{Average Linear Numbering}, shown
in Figures~\ref{fig:Ch3-Numbering-descending}
and~\ref{fig:Ch3-Numbering-linear}, respectively, for a $4 \times 4$
network.
\begin{itemize}
\item \textbf{Average Descending numbering}: node numbers increase towards the
destination and downwards,
\item \textbf{Average Linear numbering}: node numbers increase towards the
destination but vertical numbering ensures that nodes closest to
each other retain close numbering.
\end{itemize}
\subsubsection{Numbering based on instantaneous channel conditions}

\begin{itemize}
\item \textbf{Instantaneous $S-R_k$ numbering }: node numbers increase
with increasing source-relay channels. The first node has the best
source-relay channel, the second node has the second-best
source-relay channel, etc.
\item \textbf{Instantaneous $ R_k - R_m$ numbering }: nodes are
numbered to maximize the channel between adjacent nodes. The first
relay has the best source-relay channel. The second relay has the
strongest $r_1$-relay channel. Numbers are assigned in this
process to unoccupied relays. This heuristic is based on the
notion that we should maximize the capacity of each $(R_k,
R_{k+1})$ hop.
\item \textbf{Random numbering }: nodes are numbered randomly. This case evaluates
the worst-case scenario and tests the robustness of the
optimization to numbering.
\end{itemize}
\noindent These schemes are evaluated via simulations in
Section~\ref{sec:Ch3-Simulations}. As we will see, the achievable
rate is remarkably robust to the chosen numbering scheme.
\subsection{Partially Connected Network}
\label{sec:Resource-Allocation-Partially-Connected}
In this section we briefly discuss the more practical case of a
partially connected network in which some links between the nodes
in the network are unavailable. This is a generalization of the
fully-connected network discussed in
Section~\ref{sec:Resource-Allocation-Fully-Connected} above. Such
a network is more likely to represent a large scale network where,
in any case, the solution in~(\ref{eq:Ch3-final-solution}) would
be computationally infeasible.

As an example, consider the two-relay network with the link
between $r_1$ and $r_2$ is removed. The rate matrix thus becomes
\begin{eqnarray}  \label{eq:partial-matrix}
\mathbf{L}_{3} =  \left[\begin{array}{ccc} L_{s r_1} & 0 & 0 \\
L_{s r_2} & 0 & 0 \\
L_{s d} & L_{r_1 d} &  L_{r_2 d} \\
\end{array} \right].
\end{eqnarray}
Removing the link thus reduces the rank of this matrix by one, and
the rate matrix is now non-invertible, eliminating the solution
defined by $I_1 = I_2 = I_3$, where both relays are active. The
optimal solution in this case is thus to select $r_1$, $r_2$, or
not to relay. Note, however, that removing a link does not
automatically lead to a non-invertible rate matrix. Consider, for
example, the three-relay network with the link between $r_1$ and
$r_3$ removed. The corresponding rate matrix
\begin{eqnarray}  \label{eq:partial-matrix2}
\mathbf{L}_{4} =  \left[\begin{array}{cccc} L_{s r_1} & 0 & 0 & 0 \\
L_{s r_2} & L_{r_1 r_2} & 0 & 0 \\
L_{s r_3} & 0  & L_{r_2 r_3} & 0 \\
L_{s d} & L_{r_1 d} &  L_{r_2 d} & L_{r_3 d}\\
\end{array} \right]
\end{eqnarray}
\noindent is full-rank and invertible.

The approach to the optimization problem for the case of the
arbitrary connected network is that the same as for the
fully-connected network, with the exception that the rate matrix
$\mathbf{L}_{N+1}$ may not be invertible, in which case the
corresponding solution is inadmissable. The remaining steps remain
unchanged.
\section{Implementation with Reduced Complexity} \label{sec:Ch3-Implementation}

The solution to the optimization problem in~(\ref{eq:R-0}),
(\ref{eq:R-k}) and~(\ref{eq:R-T}) involves checking $2^N$
potential solutions. Although the process is conceptually simple,
each solution involves the inverse of a rate matrix. In this
section, we show how the optimization problem in the previous
section can be significantly simplified using a recursive
solution. This solution, which exploits the special structure of
the rate matrix, greatly simplifies the matrix inversion, as well
as reduces the number of possible solutions to check. Essentially,
while the solution in
Section~\ref{sec:Resource-Allocation-Fully-Connected} was a
top-down approach, the approach we suggest here is bottom-up.

Consider a set of $p$ relays, $\mathcal{P} = \{r_1, r_2, \ldots,
r_p \}, p \geq 0 $, and its corresponding rate matrix
$\mathbf{L}_{p+1}^{\mathcal{P}}$, solution vector
$\mathbf{t}_{p+1}^{\mathcal{P}}$ and maximum rate (if available)
$R^{\mathcal{P}}$. We note that if $p = 0$ and the set is empty,
the rate matrix and solution vector are constants, $L_{sd}$ and
$1$, respectively. Denote as $\mathcal{P}'$ the set $\mathcal{P}$
appended with another relay, i.e., $\mathcal{P}' = \{r_1, r_2,
\ldots, r_p, r_{p+1}\}$. Denote as
$\mathbf{L}_{p+2}^{\mathcal{P}'}$,
$\mathbf{t}_{p+2}^{\mathcal{P}'}$, and $R^{\mathcal{P}'}$
 the matrix, solution vector and rate corresponding
to set $\mathcal{P}'$.

{\bf Proposition 3:} Given   $ \left( \mathbf{L}_{p+1}^{\mathcal{P}}
\right) ^{-1}$,
 $ \left(  \mathbf{L}_{p+2}^{\mathcal{P}'}
\right)^{-1} $  can be obtained with computational complexity
order of $O(p^2)$

{\bf Proof:} For $p \geq 0$, the rate matrix $
\mathbf{L}_{p+2}^{\mathcal{P}'}$ can be written as

\begin{eqnarray}
\mathbf{L}_{p+2}^{\mathcal{P}'}  = \left[\begin{array}{ccc} \mathbf{L}_{p+1}^{\mathcal{P}}(1:p, 1:p) & \vline & \mathbf{0}_{p \times 2}  \\
\hline \mathbf{F}_{2 \times p} &  \vline & \mathbf{T}_{2}
\end{array} \right],
\end{eqnarray}
\noindent where $\mathbf{L}_{p+1}^{\mathcal{P}}(1:p, 1:p) $
denotes the first $p$ rows and columns of the rate matrix
$\mathbf{L}_{p+1}^{\mathcal{P}}$, $\mathbf{0}_{p \times 2}$ is a
$(p \times 2)$ matrix of zeros, $\mathbf{T}_{2}$ is a $(2 \times
2)$ lower- triangular matrix, and $\mathbf{F}_{2 \times p}$ is a
$(2 \times p)$ fully-loaded matrix. Note that
$\mathbf{L}_{p+1}^{\mathcal{P}}(1:p,1:p)$ is triangular. Using the
inverse of a partitioned matrix \cite{HornJohnson},
 $\left(  \mathbf{L}_{p+2}^{\mathcal{P}'} \right)^{-1}$ can be
written as
\begin{eqnarray}
\left(  \mathbf{L}_{p+2}^{\mathcal{P}'} \right)^{-1} =
\left[\begin{array}{ccc}  \left(
\mathbf{L}_{p+1}^{\mathcal{P}}(1:p, 1:p) \right) ^{-1}   &  \vline & \mathbf{0}_{p \times 2} \\
\hline
  -\mathbf{T}_{2}^{-1} \mathbf{F}_{2 \times p} \left(
\mathbf{L}_{p+1}^{\mathcal{P}}(1:p, 1:p) \right) ^{-1}   &  \vline &
\mathbf{T}_{2}^{-1}
\end{array} \right].
\end{eqnarray}
\noindent Note that $ \left( \mathbf{L}_{p+1}^{\mathcal{P}}(1:p,
1:p) \right) ^{-1} $ is the inverse of a partition of the
triangular matrix $ \mathbf{L}_{p+1}^{\mathcal{P}}$. Using the
inverse of a partitioned matrix one more time, however, it is easy
to see that
\begin{eqnarray}
 \left( \mathbf{L}_{p+1}^{\mathcal{P}}(1:p, 1:p) \right) ^{-1} =
 (\mathbf{L}_{p+1} ^{\mathcal{P}} )  ^{-1}(1:p, 1:p),
\end{eqnarray}
and thus
\begin{eqnarray} \label{eq:L-recursive}
\left(  \mathbf{L}_{p+2}^{\mathcal{P}'} \right)^{-1} =
\left[\begin{array}{ccc}
( \mathbf{L}_{p+1} ^{\mathcal{P}} )  ^{-1}(1:p, 1:p)   &  \vline & \mathbf{0}_{p \times 2} \\
\hline
  -\mathbf{T}_{2}^{-1} \mathbf{F}_{2 \times p} (\mathbf{L}_{p+1} ^{\mathcal{P}} )  ^{-1}(1:p, 1:p) &  \vline &
\mathbf{T}_{2}^{-1}
\end{array} \right],
\end{eqnarray}
and hence obtaining $\left(  \mathbf{L}_{p+2}^{\mathcal{P}'}
\right)^{-1}$ is an $O(p^2)$ operation. $\blacksquare$

Using this above proposition, the solution vector
 $\mathbf{t}_{p+2}^{\mathcal{P}'}$ of $\mathbf{L}^{\mathcal{P}'}_{p+2}$ can be
obtained from the solution vector $\mathbf{t}_{p+1}^{\mathcal{P}}$
of $\mathbf{L}^{\mathcal{P}}_{p+1}$:
\begin{eqnarray} \label{eq:time-recursive}
\mathbf{t}_{p+2}^{\mathcal{P}'}  =
\frac{ \left( \mathbf{L}^{\mathcal{P}'}_{p+2}\right)^{-1}
 \mathbf{1}_{p+2}          }   {
\mathbf{1}_{p+2}^T \left(
\mathbf{L}^{\mathcal{P}'}_{p+2}\right)^{-1} \mathbf{1}_{p+2}} =
%
%
%
%
\left[\begin{array}{c} \mathbf{t}_{p+1}^{\mathcal{P}}(1:p) \\ \hline  \mathbf{t}_{p+2}^{\mathcal{P}'}(p+1) \\
\mathbf{t}_{p+2}^{\mathcal{P}'}(p + 2)
\end{array} \right],
\end{eqnarray}
\noindent where $ \mathbf{t}_{p+1}^{\mathcal{P}}(1:p)$ represent
the first $p$ entries of the already-calculated solution vector $
\mathbf{t}_{p+1}^{\mathcal{P}}$, and
$\mathbf{t}_{p+2}^{\mathcal{P}'}(p+1)$ and
$\mathbf{t}_{p+2}^{\mathcal{P}'}(p+2)$ are the last two entries of
the solution vector $ \mathbf{t}_{p+2}^{\mathcal{P}'}$ that remain
to be calculated. $R^{\mathcal{P}'} = \frac{1}{ \mathbf{1}_{p+2}^T
\left( \mathbf{L}^{\mathcal{P}'}_{p+2}\right)^{-1}
\mathbf{1}_{p+2}}$ is the maximum achievable rate obtained using
the set $\mathcal{P}'$ of relays. The last two entries of the
solution vector $\mathbf{t}_{p+2}^{\mathcal{P}'}(p+1) $ and
$\mathbf{t}_{p+2}^{\mathcal{P}'}(p+2) $ can be written as
\begin{eqnarray} \label{eq:last-two-elements}
\left[\begin{array}{c}  \mathbf{t}_{p+2}^{\mathcal{P}'}(p+1)  \\
\mathbf{t}_{p+2}^{\mathcal{P}'}(p+2)
\end{array} \right]  =
R^{\mathcal{P}'} \left[\begin{array}{ccc}
  -\mathbf{T}_{2}^{-1} \mathbf{F}_{2 \times p} \left( \mathbf{L}^{\mathcal{P}}_{p+1}\right)^{-1}(1:p, 1:p) &  \vline &
\mathbf{T}_{2}^{-1}
\end{array} \right] \mathbf{1}_{(p+2) \times 1},
\end{eqnarray}
With a corresponding achievable rate $R^{\mathcal{P}'}$
\begin{eqnarray} \label{eq:rate-recursive}
\nonumber R^{\mathcal{P}'} &=& \frac{1}{ \mathbf{1}_{p+2}^T \left(
\mathbf{L}^{\mathcal{P}'}_{p+2}\right)^{-1}  \mathbf{1}_{p+2}} =
\left(\sum_{ij } \left(
\mathbf{L}^{\mathcal{P}'}_{p+2}\right)^{-1} (i,j) \right)^{-1},  \\
\nonumber &=&   \left( \sum_{i,j}\left(
\mathbf{L}^{\mathcal{P}}_{p+1}\right)^{-1} (i,j) - \sum_{i,j}
\mathbf{T}_{2}^{-1} \mathbf{F}_{2 \times p} \left(
\mathbf{L}^{\mathcal{P}}_{p+1}\right)^{-1}(1:p, 1:p) (i,j) +
\sum_{ij} \mathbf{T}_{2}^{-1} (i,j) \right) ^{-1} , \\
\end{eqnarray}
where we use $\sum_{i,j}\mathbf{A}(i,j)$ to denote the summation
over all the elements of matrix $\mathbf{A}$.

Using the above, the optimization problem for a network of $N$
potential relays can be solved recursively as follows:
\begin{enumerate}
\item Determine the set of all potential relay combinations.
Sequence the set as:
\begin{eqnarray}
\nonumber \mathcal{M} =  \{ (r_1), (r_1, r_2), (r_1, r_2,r_3),
\ldots (r_1, r_2, \ldots, r_N),
\\ \nonumber (r_1, r_3), (r_1, r_3,r_4),
\ldots, (r_1,
r_3, \ldots, r_N),  \\
\nonumber   \ldots \ldots
 \\ \nonumber
(r_1, r_N),
\end{eqnarray}
\begin{eqnarray}
\nonumber
(r_2), (r_2, r_3), (r_2, r_3,r_4), \ldots (r_2, r_3, \ldots, r_N),
\\ \nonumber (r_2, r_4), (r_2, r_4,r_5), \ldots, (r_2,
r_4, \ldots, r_N),  \\  \nonumber \ldots \ldots \\ \nonumber (r_2,
r_N),
%
 \\ \nonumber \ldots  \\ \nonumber(r_{N-1}, r_N) \}.
\end{eqnarray}
\noindent Note that each ``row" of $\mathcal{M}$ is a subset of
relay combinations in which each element is formed from the
previous element by adding a relay.
\item In each ``row", obtain the rate matrix, its respective
optimized time allocation vector and achievable rate for each
element (i.e., relay combination) recursively
using~(\ref{eq:L-recursive}),~(\ref{eq:time-recursive}),~(\ref{eq:last-two-elements})
and~(\ref{eq:rate-recursive}).
\item Check that for each particular set $\mathcal{P}$ of $p$ relays, the solution $\mathbf{t}_p$ and achievable rate $R_p$ satisfies the
constraints:
\begin{eqnarray}
\label{eq-constraint-1} R^{\mathcal{P}} \geq 0, \\
\label{eq-constraint-2} \mathbf{t}^{\mathcal{P}}_{p+1} >
\mathbf{0}_{p+1}.
\end{eqnarray}
\begin{itemize}
\item If both constraints are satisfied, place the solution in the
potential set of valid solutions $\mathcal{K}$, advance elements and
return to step $(1)$.
\item If~(\ref{eq-constraint-2}) is not satisfied, check which
element of the the allocation vector $\mathbf{t}_p$ does not
satisfy the constraint.
\begin{itemize}
\item  If any of the first $(p-1)$ entries of $\mathbf{t}_p$ are less than zero, i.e.,  $\mathbf{t}_p(1: p-1 <
\mathbf{0}_{p-1})$, this constraint will not be satisfied for any
other relay combinations in this ``row''. Advance rows and return to
item $(1)$.

\item If the constraint is not satisfied by either of
the last two items in the solution vector, discard the solution
but check the other elements in the ``row".
\end{itemize}
\end{itemize}
\item From the set $\mathcal{K}$, pick the highest achievable rate
and its corresponding time allocation.
\end{enumerate}
The recursive algorithm given above simplifies the optimization
problem in two ways:
\begin{enumerate}
\item It reduces the computation load of determining successive
matrix inverses by writing each matrix inverse as a function of
another, already known, matrix inverse, and two other matrices
obtained through simple matrix multiplication.
\item It may eliminate infeasible solutions by discarding relay combinations which do not satisfy constraints.
For example, if the relay combination $(r_1, r_2, r_3)$ does not
satisfy the constraints, the combination $(r_1, r_2, r_3, r_4)$ may
be automatically discarded.
\end{enumerate}
\subsection{Complexity and Number of Operations}
In the next paper we will compare relay selection schemes partly
on computational complexity. In this section we calculate this
complexity, which also quantifies the computational savings of the
recursive scheme presented above in
Section~\ref{sec:Ch3-Implementation}.

The complexity of the recursive scheme is bounded by complexity of
matrix multiplication. The number of operations (multiplications
and additions) required in the product of two matrices of size
$(m,n)$ and $(n,p)$ is $2mpn - mp$~\cite{Kaporin},
and the number of operations required for the product of a matrix
of size $(m,n)$ with a square, size-$n$ diagonal matrix is
\begin{eqnarray}
m \left( \sum_{k=0}^{n-1} k + \sum_{k=1}^{n} k     \right) = mn^2.
\label{eq:Ch3-diagonal-matrix-multiplication}
\end{eqnarray}

We now calculate the number of operations required for each rate
matrix of size $(q+1)$, corresponding to the set $\mathcal{Q}'$ of
$q$  relays. The calculation of the matrix fundamental to the
recursive algorithm,
\begin{eqnarray}
\left( \mathbf{L}^{\mathcal{Q}'}_{q+1}\right)^{-1} =
-\mathbf{T}_{2}^{-1} \mathbf{F}_{2 \times (q-1)} \left(
\mathbf{L}^{\mathcal{Q}}_{q}\right)^{-1}(1:q-1, 1:q-1).
\label{eq:Ch3-matrix-multiplication}
\end{eqnarray}
requires a total of
%
$2q^2 + 2q + 1$
 operations, broken down as:
\begin{enumerate}
\item $ -\mathbf{T}_{2}^{-1} \rightarrow 5$ operations,
\item $-\mathbf{T}_{2}^{-1}
\mathbf{F}_{2 \times (q-1)} = \mathbf{A}_{2 \times (q-1)}
\rightarrow 6(q-1)$ operations using $2mpn - mp$,
\item $\mathbf{A}_{2 \times (q-1)} \left( \mathbf{L}^{\mathcal{Q}}_{q}\right)^{-1} \left( 1:q-1,
1:q-1 \right)  \rightarrow 2(q-1)^2$ operations,
using~(\ref{eq:Ch3-diagonal-matrix-multiplication}).
\end{enumerate}
From~(\ref{eq:rate-recursive}), the number of operations required
to calculate $R^{\mathcal{Q}'}$ is
%
%
$q^2 + 2q + 4$.
%
%
Using~({\ref{eq:last-two-elements}}), the number of operations
required to update the solution vector is
%
 $1 + 2(q+1) = 2q + 3$.
%
%
Summing
the above, we obtain the total number of operations required in
one iteration of the resource allocation algorithm:
\begin{eqnarray}
\mathrm{Op}(q) = (2q^2 + 2q + 1)  +
 (q^2 + 2q + 4) + (2q + 3) = 3q^2 + 6q + 8.
 \label{eq:Ch3-Op}
\end{eqnarray}
Note that the complexity order of calculating each rate and
solution vector is $O(q^2)$. Without the recursion, this
complexity is of order $O(q^3)$, resulting from the inverse of the
rate matrix. The recursion thus introduces significant savings in
terms of complexity.

We now calculate the worst-case total number of operations
required by the resource allocation algorithm. In the worst case,
the algorithm cycles through $2^N$ operations consisting of ${N
\choose q}$ sets of $q$ relays which require $3q^2 + 6q + 8$
operations. The total worst case number of operations is therefore
\begin{eqnarray}
\sum_{q = 1}^ {N}   {N \choose q} (3q^2 + 6q + 8 ).
\label{eq:Ch3-worst-case-number-operations}
\end{eqnarray}
This calculation could be rendered more precise if it were
possible to account for the savings obtained in
Section~\ref{sec:Ch3-Implementation} which eliminates some
infeasible solutions a priori by discarding relay combinations
known to not satisfy the constraints. The probability of this
occurring for particular channel realizations is unfortunately
very difficult to compute, and we thus show only the worst-case
result.
\section{Simulations} \label{sec:Ch3-Simulations}
In this section, we present results of the resource allocation
scheme discussed in Section~\ref{sec:Resource-Allocation} for
networks with $1$ to $6$ relays arranged linearly, and $4$ and $9$
nodes arranged in a grid. The figure of merit is the achievable
rate $R_a$ with an outage probability of $10^{-3}$, i.e.
$\mathrm{Pr}[R^* < R_a] = 10^{-3}$. A closed form expression for
the outage probability of optimized cooperation is very
complicated and beyond the scope of the paper. The outage
probability and rate are thus obtained numerically.

The relays are equispaced on a line between the source and
destination, as in Figure~\ref{fig:Ch3-Relays-Line}, and we use an
attenuation exponent of $p_a = 2.5$. This choice is motivated by
the application of static mesh-nodes installed on posts;
transmissions between such nodes should undergo little shadowing
and a lower attenuation exponent. From $60000$ fading realizations
we obtain the cumulative density function of the instantaneous
rate $F_R(r)$. The outage rate is the rate for which the
probability of outage is $10^{-3}$, i.e., $F^{-1}_R(10^{-3})$.

\begin{figure}
\begin{center}
\includegraphics[width=4.5in]{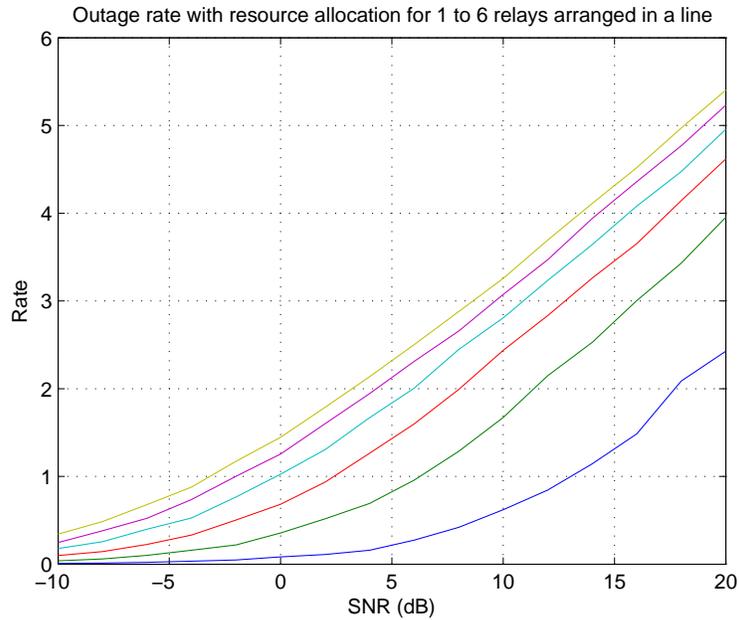}
\caption{Outage rate vs. SNR using $1, \ldots 6$ potential relays
and with resource allocation.} \label{fig:Ch3-Rate-RA}
\end{center}
\end{figure}

\begin{figure}
\begin{center}
\includegraphics[width=4.5in]{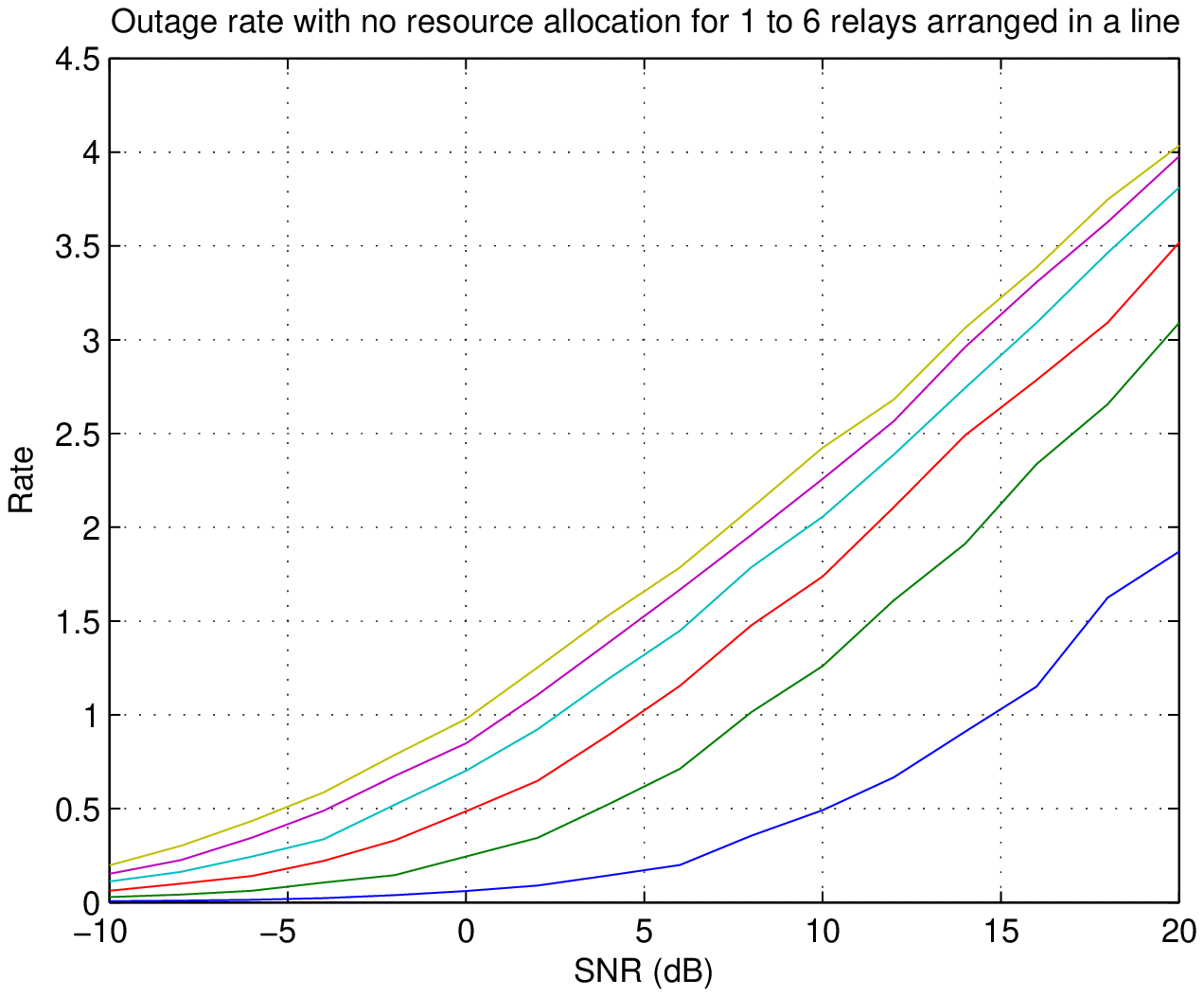}
\caption{Outage rate vs. SNR using $1, \ldots 6$ potential relays
and without resource allocation.} \label{fig:Ch3-Rate-NRA}
\end{center}
\end{figure}

Figure~\ref{fig:Ch3-Rate-RA} and~\ref{fig:Ch3-Rate-NRA} plot the
outage rate as a function of the average end-to-end SNR,
$\frac{P}{N_0 W}$, for optimized and non-optimized cooperation,
respectively. The rate for the optimized cooperation is obtained
from~(\ref{eq:Ch3-final-solution}). Non-optimized cooperation uses
equal time allocation, i.e., the rate for a particular relay set
is simply the minimum of the mutual information at each node.
Non-optimized cooperation, however, does optimally select relays
by choosing the best, in terms of rate, of the $2^N$ relay
combinations.
Comparing Figure~\ref{fig:Ch3-Rate-RA} and
Figure~\ref{fig:Ch3-Rate-NRA} shows that optimizing resources
increases rates significantly, as expected. The outage rate
increases as a function of nodes available to relay. We also note
the typical phenomenon of decreasing marginal returns: the gains of
adding each additional relay decreases with increasing number of
relays.

\begin{figure}
\begin{center}
\includegraphics[width=4.5in]{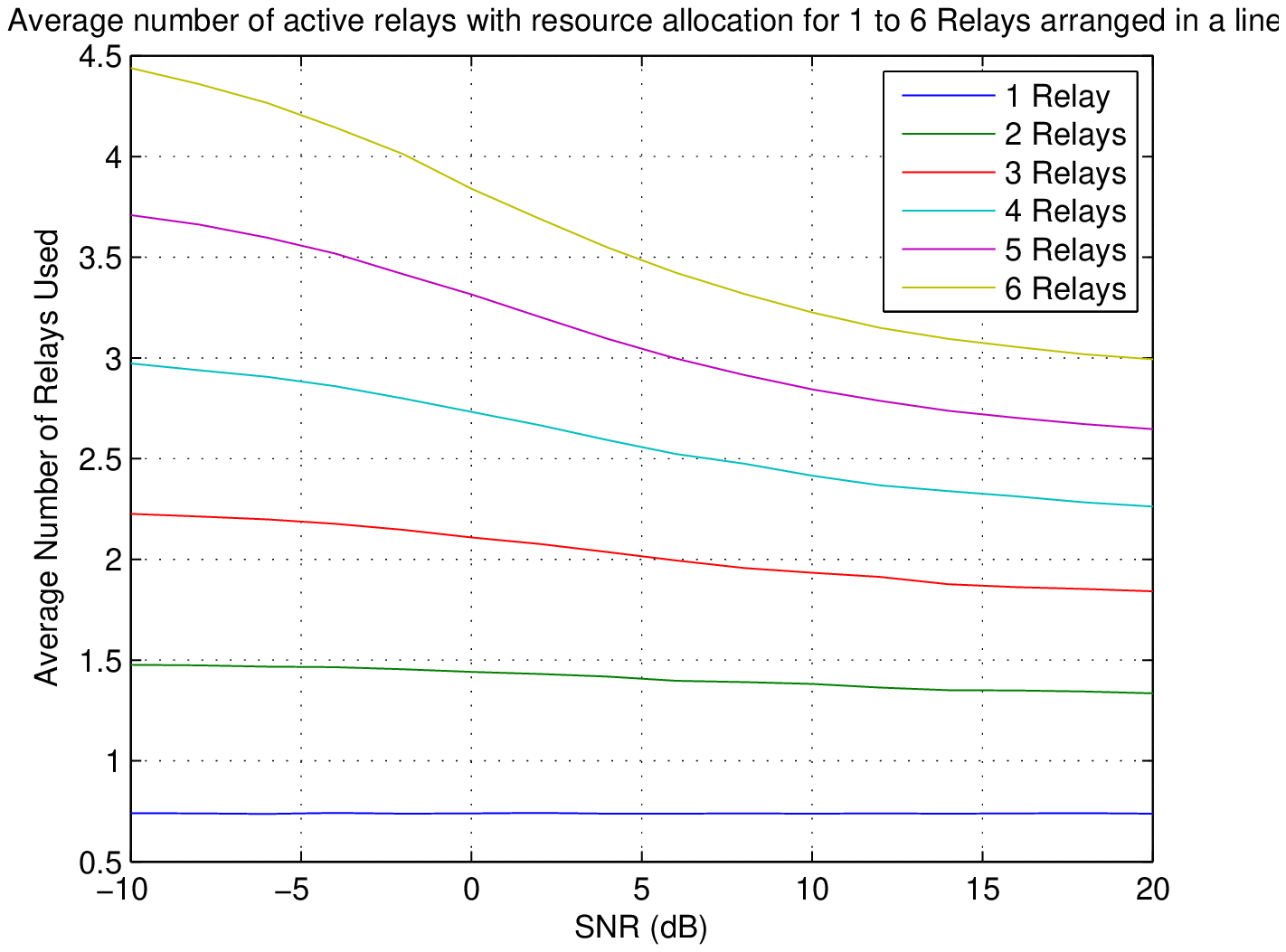}
\caption{Average number of active relays with $1, \ldots 6$
potential relays and with resource allocation.}
\label{fig:Ch3-Number-RA}
\end{center}
%
%
%
\begin{center}
\includegraphics[width=4.5in]{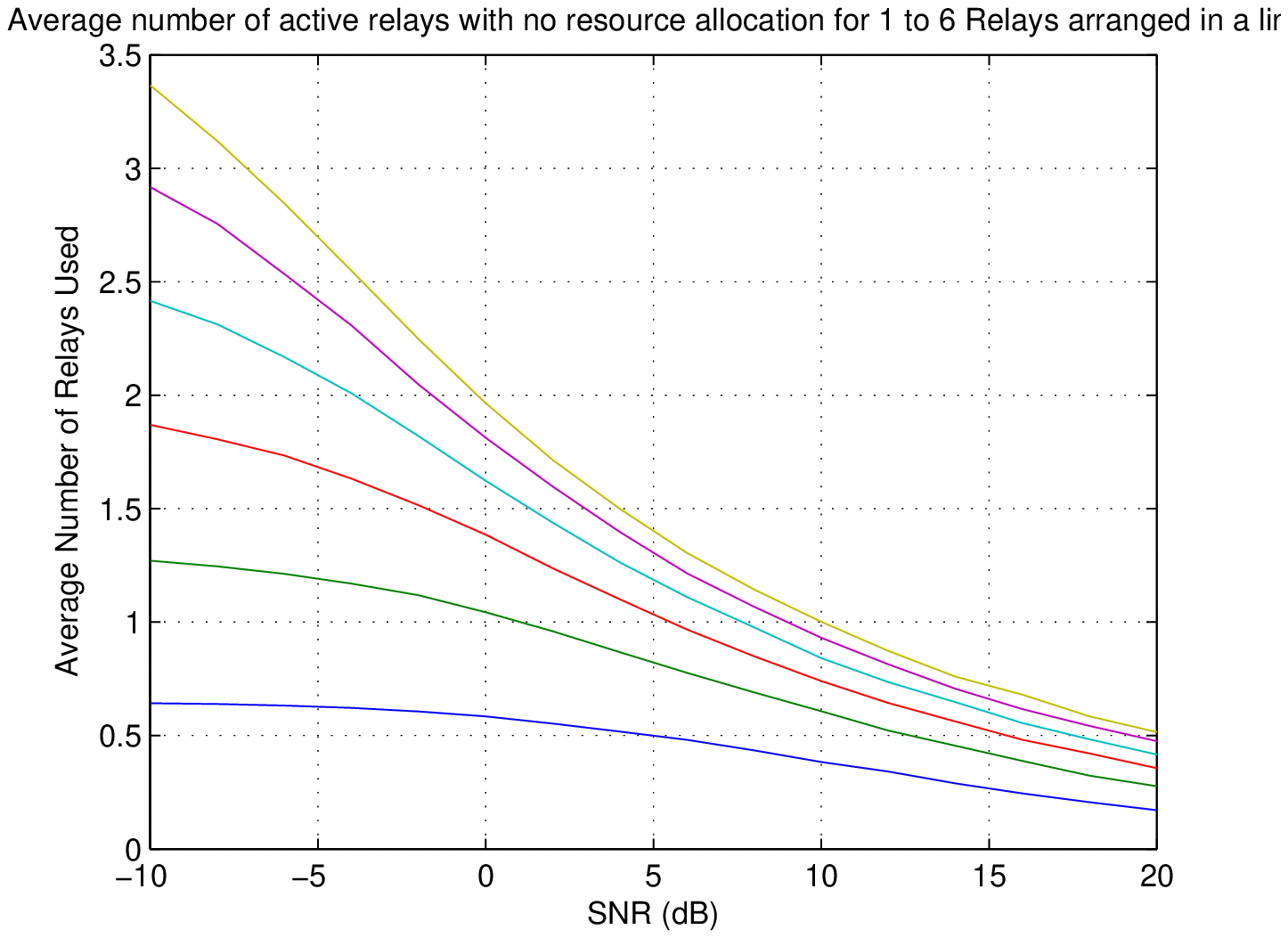}
\caption{Average number of active relays with $1, \ldots 6$
potential relays and without resource allocation.}
\label{fig:Ch3-Number-NRA}
\end{center}
\end{figure}

Figures~\ref{fig:Ch3-Number-RA} and~\ref{fig:Ch3-Number-NRA} show
the average number of relays that are active from the set of
potential relays for optimized and non-optimized cooperation. For
each network size, this number is a decreasing function.
Interestingly, the number of active relays decreases much faster for
non-optimized as compared to optimized cooperation, suggesting that
optimizing resources distributes the relaying burden more
effectively.

\begin{figure}
\begin{center}
\includegraphics[width=4.5in]{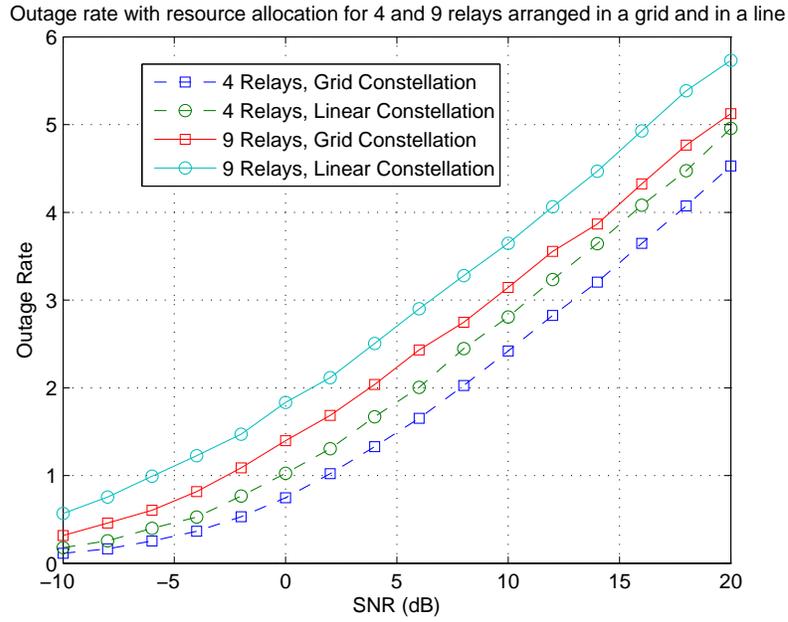}
\caption{Outage rate vs. SNR using resource allocation and for $4$
and $9$ relays arranged in a grid and in a line. }
\label{fig:Ch3-Rate-Line-Grid}
\end{center}
\end{figure}

To test the effect of geometry on the outage rate, we compare the
rates obtained by optimizing resources and the placing relays on a
line, as in Figure~\ref{fig:Ch3-Rate-RA} to those obtained by
placing the relays on a regular square grid. We number the relays
in the grid in ascending order downwards and towards the source; a
derivation of the optimal numbering is beyond the scope of this
paper. The results are demonstrated in
Figure~\ref{fig:Ch3-Rate-Line-Grid}, where we place $4$ and $9$
relays on a $2 \times 2$ and $3 \times 3$ square grid. As shown in
the figure, the rate for the linear constellation is significantly
higher than that obtained by the grid constellation, suggesting
that the path-loss incurred by traversing all the nodes laterally
results in non-negligible performance loss.

\begin{figure}
\begin{center}
\includegraphics[width=4.0in]{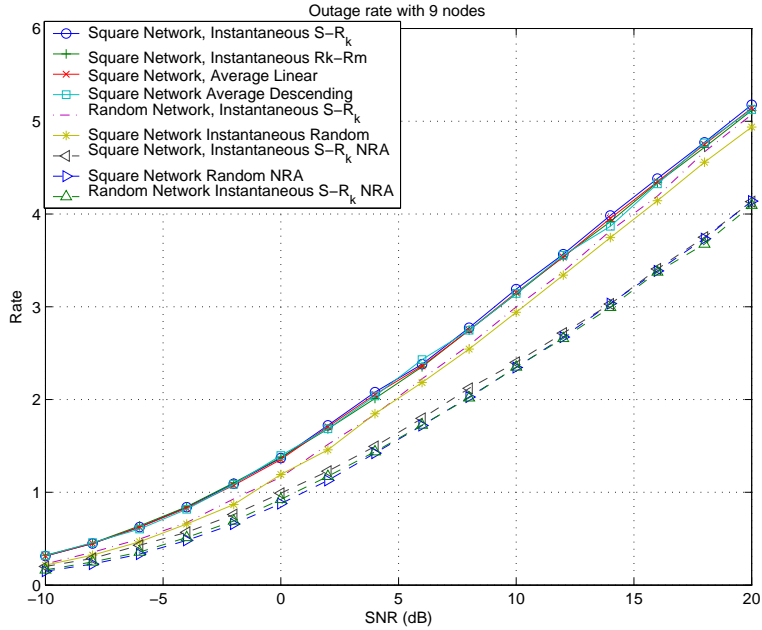}
\caption{Outage rate vs. SNR using resource allocation and for
various numbering schemes for $9$ potential nodes arranged in a grid
} \label{fig:Ch3-Numbering-Rate}
\end{center}
\end{figure}

We evaluate the performance of the numbering schemes discussed in
Section~\ref{Ch3:Numbering} in
Figure~\ref{fig:Ch3-Numbering-Rate}. The four schemes, including
two based on average channel conditions and two based on
instantaneous channel conditions, exhibit indistinguishable
performance in terms of rate. There is an expected drop in rate
with random numbering, though we note that this drop is no more
than approximately $0.25$ bits/channel use. The algorithm is thus
quite robust to numbering schemes.

Figure~\ref{fig:Ch3-Numbering-Rate} also shows the outage rate for
a network with randomly placed nodes. Here the node locations are
chosen from a uniform distribution over an area equivalent to that
of the square grid . The internode channels are obtained as
before. This example eliminates possible dependencies of the
results obtained earlier on the chosen array geometry. The
numbering here is based on the instantaneous $S-R_k$ channels. In
such a random network, as expected, the available outage rate is
lower than in a square grid network; however, at higher SNR levels
this difference disappears. Again, the significant gains due to
resource allocation are clear.

\begin{figure}
\begin{center}
\includegraphics[width=4.0in]{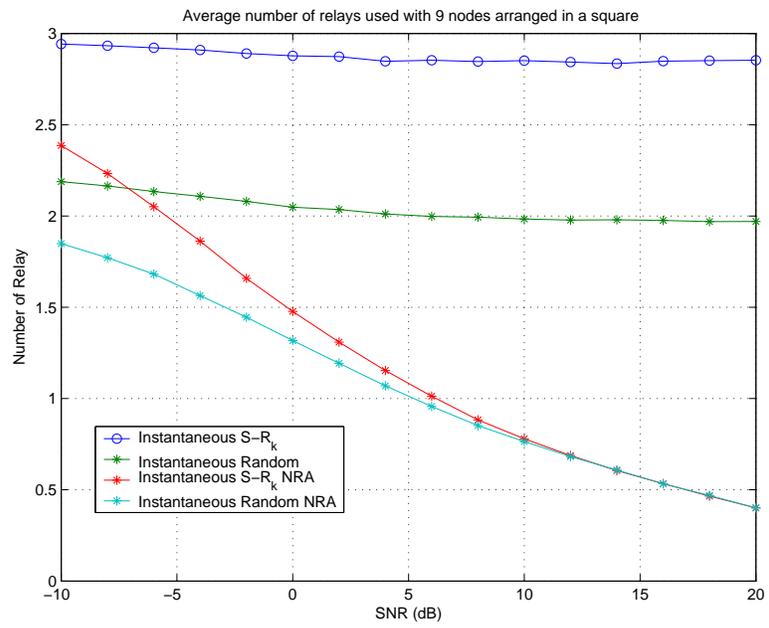}
\caption{Average number of active relays using resource allocation
and with various numbering schemes for $9$ potential nodes arranged
in a grid  } \label{fig:Ch3-Numbering-Relays}
\end{center}
\end{figure}

In Figure~\ref{fig:Ch3-Numbering-Rate} we also compare the effect of
numbering when used without resource allocation, and show only the
case of instantaneous $S-R_k$ numbering and random numbering. The
improvement from instantaneous over random numbering in this case is
less than $0.1$ bits/channel use. The robustness of the numbering
scheme thus increases by eliminating time optimization. To gain
insight into this phenomenon, in
Figure~\ref{fig:Ch3-Numbering-Relays} we plot the average number of
active users for the instantaneous and random numbering schemes with
and without resource allocation. We first observe that the
instantaneous numbering scheme uses more relays than the random
numbering scheme when resource allocation is used, and that this
difference is constant over the SNR region of interest. Without
resource allocation, on the other hand, the number of relays used
when using instantaneous and random numbering decreases quickly and
is constant for SNR values higher than $10$ dB. It is clear from
this figure that the difference in rate performance between
instantaneous and random numbering is an increasing function of the
number of selected relays. Because so few relays are selected
without resource allocation, the effect of the numbering scheme is
negligible. The influence of the numbering scheme increases when
time allocation is introduced, increasing the number of relays used
for both numbering schemes and increasing the sensitivity to the
numbering scheme. This sensitivity increases slowly, however, and is
negligible for the various numbering schemes based on heuristics.
\section{Conclusions} \label{sec:Conclusions}
In this paper, we determined the optimal channel resource
allocation, in terms of time allocation,  for the $N$-node
cooperative diversity, multihop network using DF and independent
codebooks. For a particular network, i.e., set of potential
relays, the unique solution for a particular relay numbering
scheme is obtained by taking the inverse of the triangular rate
matrix, and the optimal solution is found by maximizing over the
rate for each possible network, given its maximum size. Through
simulations, however, the optimization is shown to be robust to
the numbering scheme. We show that by exploiting the special
structure of the rate matrix, the optimization can be performed in
a recursive fashion which decreases the computation load of the
rate matrix inverse and the number of required iterations. Node
selection is inherent to the optimization strategy. Simulation
results show significant gains in achievable rate due to resource
allocation, but diminishing marginal returns as a function of
network size. Furthermore, we show a significant benefit to
arranging the nodes in a linear, as opposed to a grid,
constellation.


\bibliographystyle{IEEEtran}
\bibliography{IEEEabrv,references}

\newpage


\end{document}